 \documentclass[final,3p,times,sort&compress]{elsarticle}
 \usepackage{graphicx,amssymb,amsthm,amsmath,lineno,natbib}
 \usepackage{dcolumn,multirow}
 \usepackage[center]{subfigure}
 \usepackage{threeparttable}
 \usepackage{enumitem}
 \usepackage[mathscr]{euscript}
 \usepackage{mathtools, cuted}
 \usepackage[english]{babel}
 \usepackage{lipsum}
 
\biboptions{square}

\begin{document}

\begin{frontmatter}



\title{Critical behavior of the stochastic SIR model on random bond-diluted lattices}


\author[ufsa]{Carlos Handrey A. Ferraz\corref{cor1}}
\ead{handrey@ufersa.edu.br}
\author[ufsa]{Jos\'e Luiz S. Lima}
\cortext[cor1]{Corresponding author}
\address[ufsa]{Exact and Natural Sciences Center, Universidade Federal Rural do Semi-\'Arido-UFERSA, PO Box 0137, CEP 59625-900, Mossor\'o, RN, Brazil}

\begin{abstract}

In this paper, we investigate the impact of bond-dilution disorder on the critical behavior of the stochastic SIR model. Monte Carlo simulations were conducted using square lattices with first- and second-nearest neighbor interactions. Quenched bond-diluted lattice disorder was introduced into the systems, allowing them to evolve over time. By employing percolation theory and finite-size scaling analysis, we estimate both the critical threshold and leading critical exponent ratios of the model for different bond-dilution rates ($p$). An examination of the average size of the percolating cluster and the size distribution of non-percolating clusters of recovered individuals was performed to ascertain the universality class of the model. The simulation results strongly indicate that the present model belongs to a new universality class distinct from that of 2D dynamical percolation, depending on the specific $p$ value under consideration.

\end{abstract}


\end{frontmatter}


\section{Introduction \label{sec:int}}

In epidemiology, understanding the spread of infectious diseases is crucial for disease control and public health management. Bond-diluted lattices offer a platform to study the impact of network structure on disease transmission dynamics. By examining how diseases propagate on networks with partial connections, it is possible to develop more effective strategies for disease prevention, intervention, and control. A fundamental system of ordinary differential equations was initially developed for modeling the dynamics of infectious diseases within populations \cite{Kermack1927}. This system categorized individuals into three basic classes: susceptible (S), infected (I), and recovered (R). This gave rise to the SIR model, named after its component classes. Over time, this model has undergone numerous adaptations and extensions, such as the SIS, SIRS and SEIR models, each tailored to capture different nuances of disease transmission within populations.

The simplicity of SIR model lies in its assumption that infected individuals either gain permanent immunity or succumb to the disease, thereby exiting the transmission dynamics. This characteristic renders the model particularly suitable for simulating outbreaks of influenza\cite{Kim2020}, measles \cite{Kassem2010, Tilahun2020}, varicella \cite{Sun2023}, AIDS \cite{Huang2011, Bashir2017}, COVID-19 \cite{Atkeson2020, Yang2021, Gounane2021}, and many others. This model, rooted in both collective dynamics \cite{Wang2016, Shao2022} and complex systems \cite{Helbing2015}, serve as invaluable tool for assessing disease propagation rates and evaluating the efficacy of public health interventions. It has been shown that the SIR model shares a universality class with dynamic percolation \cite{Grassberger1983, Munoz1999, Ziff2010, Souza2011, Pastor2015}.

Lattice models excel in capturing various aspects of infection spread across spatially extended landscapes, where wave-like evolution plays a key role. This work focuses on a stochastic lattice-gas variant of the SIR model with asynchronous site updates, a framework previously employed in analyzing various population dynamics \cite{Satulovsky1994, Antal2001}. In both synchronous and asynchronous implementations, the model exhibits a phase transition as control parameters vary. This transition, typically second-order, delineates two distinct regimes: an endemic phase where the population largely remains susceptible, and an epidemic phase characterized by widespread disease transmission, leading to a significant proportion of the population becoming infected, recovering, or succumbing to the disease. At the critical transition point, the system reaches an epidemic outbreak threshold. It is worth mentioning that both stochastic and deterministic descriptions of the model can be connected through mean-field approximation \cite{Lugo2008, Tome2010}, resulting in Langevin equations associated with the Fokker--Planck equation. 

In the present study, we will conduct an analysis focusing on two key aspects: the average size of the percolating cluster and the size distribution of non-percolating clusters among recovered individuals. To achieve this, we will employ the Newman--Ziff algorithm \cite{Newman2001}, which facilitates the determination of critical exponent ratios within the model.

The SIR stochastic lattice model behaves akin to an absorbing system, characterized by an active phase featuring an infinite array of absorbing configurations wherein the final state solely comprises recovered individuals. However, our approach differs: we halt simulations upon confirming the existence of a percolating cluster (spanning cluster) within the system, effectively establishing a non-absorbing state for the model. This strategy expedites data analysis without compromising the assurance of reaching the system's asymptotic limit.

Notably, in the critical regime of the system, only a single percolation cluster is generated. Through a meticulous analysis of these clusters, we aim to ascertain the universality class of the model. Furthermore, the network topology serves as a crucial determinant directly influencing the dynamics of involved processes \cite{Ferraz2008, Ferraz2017, Ferraz2018}. Therefore, we will investigate how quenched bond-diluted lattice disorder influence both dynamics and universality class of this model.

In this context, bond-diluted lattices emerge from regular square lattices ($N=L\times L$) initially featuring both first- and second-nearest neighbor interactions. In these lattices, each site is subject to a probability $p$ of experiencing disconnection in its connections (dilution), with the requirement that at least one connection remains intact, thereby preventing isolated sites. These networks aim to emulate a more realistic population, reflecting a diverse array of interconnected relationships among individuals. The present work investigates several cases of varying $p$.

This study mainly aims to understand how the dilution factor affects the critical behavior of the stochastic SIR model. Since the dilution probability $p$ can be understood as a quenched topological disorder, we want to determine if this kind of disorder is relevant to changing the universality class of the model. Several criteria have been devised to anticipate whether quenched disorder might alter the critical exponents of a given model, including the well-known Harris criterion \cite{Harris1974} and its refinement, the Harris--Barghathi--Vojta criterion (HBV) \cite{Barghathi2014}. The Harris criterion posits that a second-order transition in a $d$-dimensional system, with an original correlation length exponent $\nu$, remains stable against quenched spatial disorder if $\nu > 2/d$. Conversely, the HBV criterion suggests that quenched topological disorder is inconsequential concerning phase transition stability if the system satisfies $\nu > 1/k$, where $k$ denotes the disorder decay exponent measuring how swiftly coordination number fluctuations decrease with increasing system length scale. However, it is worth noting that these criteria have been observed to falter in certain cases \cite{Schrauth2018}.

The manuscript is organized as follows: In Section \ref{sec:sir}, we provide an overview of the SIR stochastic lattice model. Section \ref{sec:mcs} details the Monte Carlo (MC) simulation background and finite-size analysis. In Section \ref{sec:r}, we present and discuss the results. Finally, Section \ref{sec:c} we present our conclusions.

\section{SIR stochastic lattice model\label{sec:sir}}

The SIR stochastic lattice model is formulated on a lattice consisting of $N$ sites, where each site can host a single individual characterized as susceptible ($S$), infected ($I$), or recovered (immunized/deceased) ($R$). This model encompasses two distinct processes: an auto-catalytic mechanism, denoted by $I+S\rightarrow I+I$, and a spontaneous transition from infection to recovery, represented by $I\rightarrow R$. At each time step, a site is randomly chosen, and a specific set of dynamic rules is applied in the following way:

\begin{enumerate}[label=(\roman*)]
  \item If the site is in the state $S$ and there is at least one neighboring site in the sate $I$ then the site becomes $I$ with probability proportional to a parameter $\mu$ and the number $z$ of neighboring sites, i.e., $ \mu{\kern 1pt} m /z$, where $m$ is the number of neighboring $I$ sites.
  \item If the site is in state $I$ it becomes $R$ spontaneously with probability $\lambda$.\\
  \item If the site is $R$ it remains unchanged.
\end{enumerate}

At each site $i$ of a lattice we assign a stochastic variable $\sigma_{i}$ that can take the values 0, 1 or 2, according to whether the site is in the state $S$, $I$, or $R$, respectively. Since the transitions between states in this model are non-equilibrium ones, the allowed transitions of the state $i$ of a site are cyclic, this is, $0 \rightarrow 1 \rightarrow 2$. The corresponding transition rate is represented by $w_{i}(\sigma)$ and describes the transition $\sigma \rightarrow \sigma '$ in which the whole microscopic configuration (microstate) $\sigma ' \equiv (\sigma _1 ,\ldots ,\sigma ' _i ,\ldots ,\sigma _N )$ differs from $\sigma$ only by the state of the $i$-th site. It is given by
\begin{equation} \label{eq:1}
	w_i (\sigma ) = \frac{\mu}
{z}\delta (\sigma _i ,0)\sum\limits_j {\delta (\sigma _j ,1)}  + \lambda{\kern 1pt} \delta (\sigma _i ,1),
\end{equation}
where the summation runs over the nearest neighbors of site $i$ and $\delta(x,y)$ denotes the Kronecker delta. The parameters $\mu$ and $\lambda$ are related to the subprocesses above described, and are chosen such that $\mu+\lambda=1$. 

The system evolves in time according to a master equation for the probability distribution $W(\sigma,t)$ described by
\begin{equation} \label{eq:2}
	\frac{d}{{dt}}W(\sigma ,t) = \sum\limits_i {\{ w_i (\bar \sigma )} W(\bar \sigma ,t) - w_i (\sigma )W(\sigma ,t)\},
\end{equation}
where the microstate $\bar \sigma$ is obtained from $\sigma$ by an anticyclic permutation ($2 \rightarrow 1 \rightarrow 0$) of the state of the site $i$.

\section{ Numerical simulation and finite-size scaling analysis\label{sec:mcs}}

We can establish an asynchronous, non-absorbing SIR model by applying the following kinetic Monte Carlo rules:

\begin{enumerate}[label=(\roman*)]
	\item First, we start with a single central infected site (seed) and the remaining ones being all susceptible on a two-dimensional lattice in which each individual of the population is attached to its respective lattice site. In order to speed up the simulation we create two lists that are updated at each algorithm step: a list of infected individuals (infected list) and a list of recovered individuals (recovered list), which begins empty. 
	\item Next, we update the system state by randomly choosing an available infected site from the infected list and proceed as follows:
	\begin{enumerate}
	\item Generate a random number ($rn$) in the interval $(0,1)$. If $rn\leq \lambda$, the infected site is removed from the infected list and placed in the recovered list;
		\item Otherwise (if $rn>\lambda$), pick randomly one nearest neighbor of the infected site and make it also infected provided that it is susceptible, adding it to infected list.
	\end{enumerate}
	\item Repeat asynchronously the step (ii) several times until either there is no infected sites (endemic phase) or there is a percolating cluster of recovered sites (non-absorbing epidemic phase).
\end{enumerate}

Remarkably, research has shown that the SIR model on square lattices shares the same universality class as dynamic percolation (DP). This enables us to explore the phase transition occurring in the current non-absorbing SIR model using percolation theory. Consequently, we can define the epidemic phase of the model when a percolating cluster of recovery sites forms in the system and the endemic phase when it does not. Such a graphical analogy has also been applied to other compartmental models.

Following classical percolation theory, it is crucial to first determine the cluster distribution of recovery sites, i.e., the number of clusters with $s$ recovery sites denoted as $n_{p}(s)$. This can be achieved using the Newman--Ziff algorithm, which also features a built-in capability to identify whether a percolating cluster has formed or not based on the considered $\lambda$ value. It's noteworthy that for systems with non-periodic boundaries, like the ones under consideration here, the percolating cluster is essentially a spanning cluster \cite{Newman2001, Sen2011}.

\begin{table*}[!t] \small
\caption{\label{tab:1} Estimates of the epidemic threshold $\lambda_{c}$ and critical exponents ratios for each $p$ cases.}
\centering
\begin{tabular}{ccccc}
\hline 
 & & & & \\
 Lattice &  Epidemic threshold & $1/\nu$ & $\beta/\nu$ & $\gamma/\nu$  \\ \\ \hline
 & & & &  \\
$p=0$&$\lambda _{c}=0.275(0)$&$0.716\pm 0.014$&$0.112\pm 0.003$&$1.776\pm 0.009$ \\
$p=0.10$&$\lambda _{c}=0.266(3)$&$0.667\pm 0.022$&$0.048\pm 0.006$&$1.722\pm 0.009$ \\
$p=0.20$&$\lambda _{c}=0.258(1)$&$0.691\pm 0.019$&$0.054\pm 0.005$&$1.744\pm 0.012$ \\
$p=0.30$&$\lambda _{c}=0.250(5)$&$0.788\pm 0.011$&$0.144\pm 0.002$&$1.827\pm 0.004$ \\
$p=0.40$&$\lambda _{c}=0.241(0)$&$0.883\pm 0.013$&$0.208\pm 0.005$&$1.856\pm 0.005$ \\
$p=0.50$&$\lambda _{c}=0.226(7)$&$0.743\pm 0.009$&$0.076\pm 0.002$&$1.761\pm 0.004$ \\
$p=0.60$&$\lambda _{c}=0.213(3)$&$0.775\pm 0.018$&$0.100\pm 0.004$&$1.780\pm 0.005$ \\
$p=0.70$&$\lambda _{c}=0.196(7)$&$0.790\pm 0.016$&$0.124\pm 0.004$&$1.795\pm 0.005$ \\

DP &$-$&$3/4$&$5/48$&$43/24$ \\
\hline 
 & & & &
\end{tabular}
\end{table*}

It is worth mentioning that MC time $t$ could be determined by incrementing $t$ by $\delta t=1/n_{I}$, where $n_{I}$ represents the current number of infected sites, each time an infected site is picked from the list. However, in this context, we do not track time as our primary focus lies on static quantities such as the fraction and mean cluster size of $R$ sites.

From the cluster size distribution, we have the fraction of recovery sites in the finite (non-percolating) cluster with $s$ size
\begin{equation} \label{eq:3}
	P_s  = s\frac{{n_p (s)}}{{n_R }},
\end{equation}
 where $n_R$ is the total number of recovery sites and $n_{p}(s)$ is the number of clusters with $s$ recovery sites. Furthermore, the fraction of recovery sites in the percolating cluster $P_\infty$ can be obtained by
\begin{equation} \label{eq:4}
	P_\infty   = 1 - \frac{1}{{n_R }}\sum\limits_s {s{\kern 1pt} n_p (s)},
\end{equation}
such that the above summation excludes the percolation cluster. Now we can define the order parameter from Eq.~(\ref{eq:4}) as 
\begin{equation} \label{eq:5}
	P =  < P_\infty >, 
\end{equation}
where $<x>$ means an average taken over different dynamic realizations. The epidemic phase of the model is reached when $P\neq 0$, that is, when the percolating cluster density is non-zero; while the endemic phase happens when $P=0$. Other important quantities are the mean cluster size
\begin{equation} \label{eq:6}
	S = \frac{1}{{n_R }}\sum\limits_s {s^2 {\kern 1pt} n_p (s)},
\end{equation}
which plays the rule of the susceptibility in classical percolation theory \cite{Kirkpatrick1971,Stauffer2014} when taking the average over different runs, i.e.,
\begin{equation} \label{eq:7}
	\chi=<S>,
\end{equation}
the overall mean cluster size
\begin{equation} \label{eq:8}
	S' = \frac{1}{{n_R }}\sideset{}{'}\sum_{s} {s^2 {\kern 1pt} n_p (s)}, 
\end{equation}
and the mean quadratic cluster size
\begin{equation} \label{eq:9}
	M' = \frac{1}{{n_R }}\sideset{}{'}\sum_{s} {s^3 {\kern 1pt} n_p (s)}, 
\end{equation}
where the primed summations above also include the percolating cluster. It is worth remarking that the last two quantities $S'$ and $M'$ only make sense for finite lattice as in the asymptotic limit $(N\rightarrow \infty)$, the percolating cluster size diverges. 

\begin{figure*}[!t]
\centering
\begin{minipage}[t]{1.0\linewidth}
\centering
\subfigure[]{\label{fig:01a}\includegraphics[scale=0.40, angle=0]{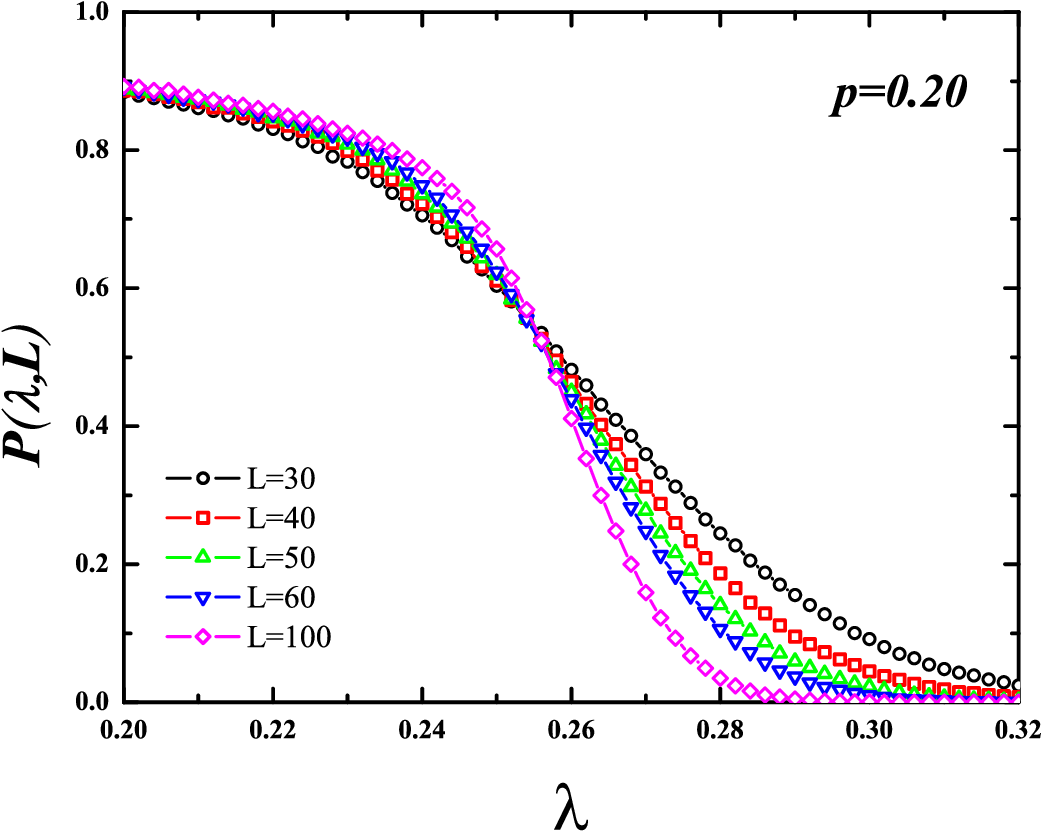}}
\qquad
\subfigure[]{\label{fig:01b}\includegraphics[scale=0.40, angle=0]{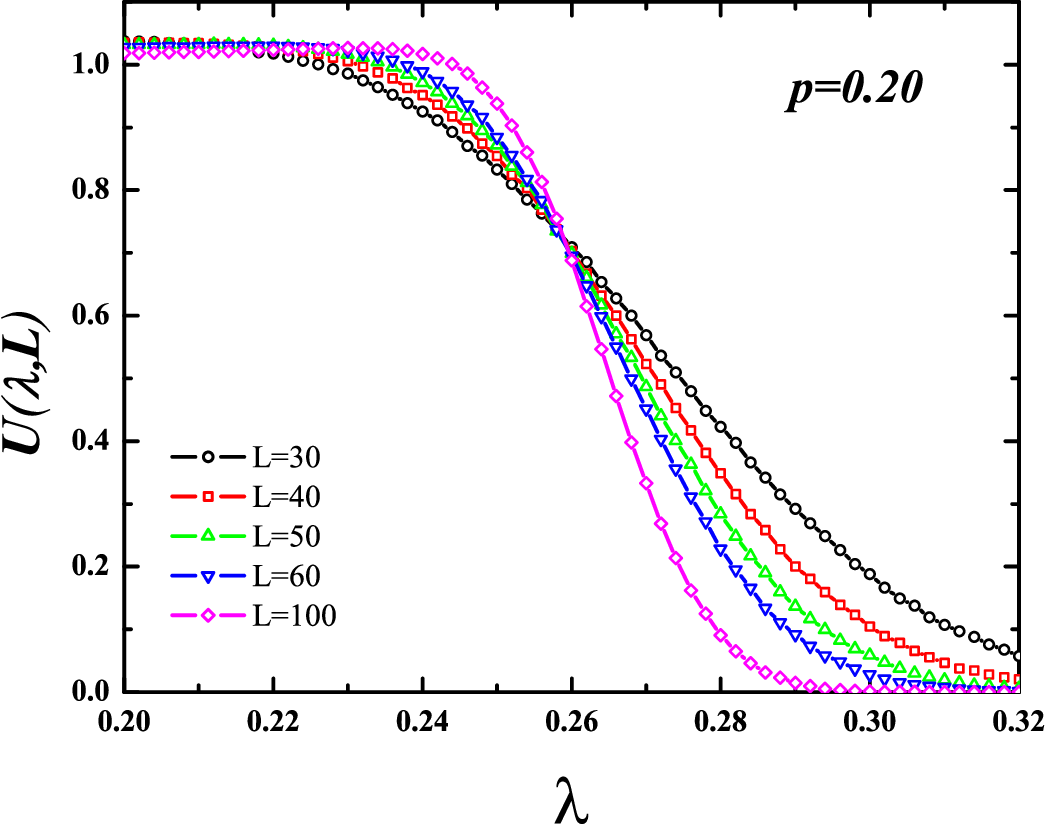}} 
\subfigure[]{\label{fig:01c}\includegraphics[scale=0.40, angle=0]{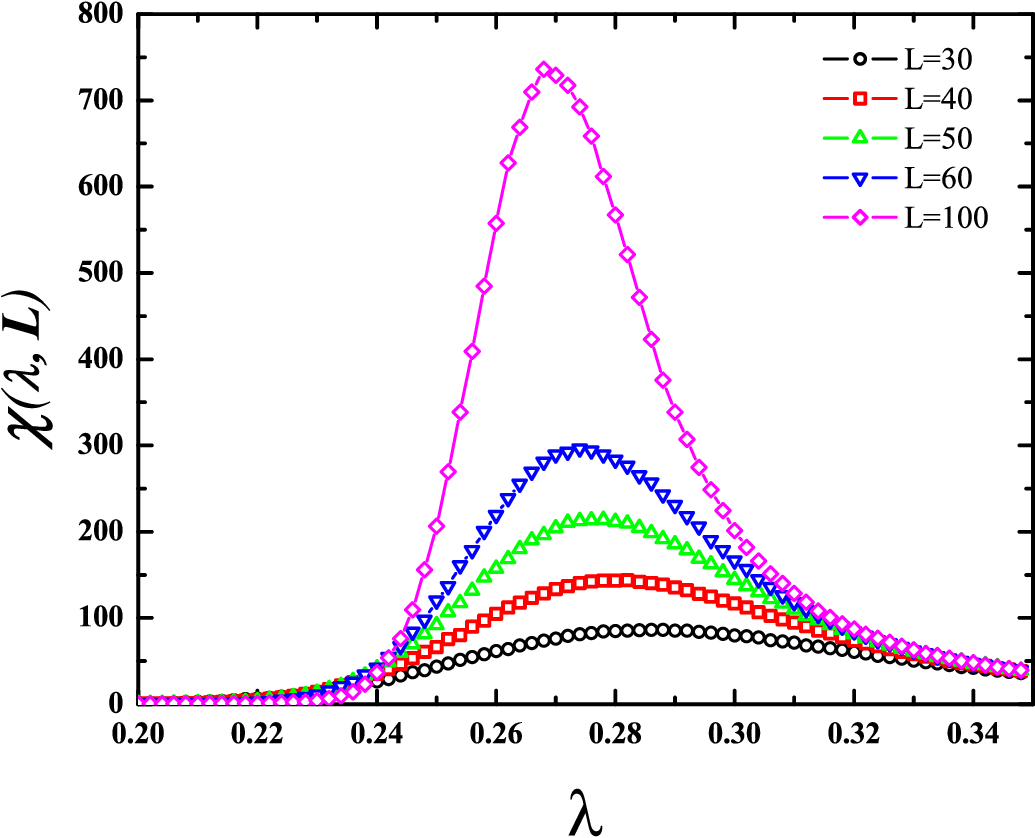}} 
\end{minipage}
\caption{Static quantities for the stochastic SIR model on square lattices with first- and second-neighbor interactions by assuming a dilution rate of $p=0.20$. Panels (a), (b), and (c) display the order parameter $P$, Binder cumulant $U$, and susceptibility $\chi$ as a function of the recovery rate $\lambda$, respectively. From the Binder cumulant crossing, we can estimate the epidemic threshold at $\lambda _{c}=0.258(1)$ for this case. Each data point represents averages over $10^5$ independent realizations.}\label{fig:01}
\end{figure*}

\begin{figure*}[!t]
\centering
\begin{minipage}[t]{1.0\linewidth}
\centering
\subfigure[]{\label{fig:02a}\includegraphics[scale=0.40, angle=0]{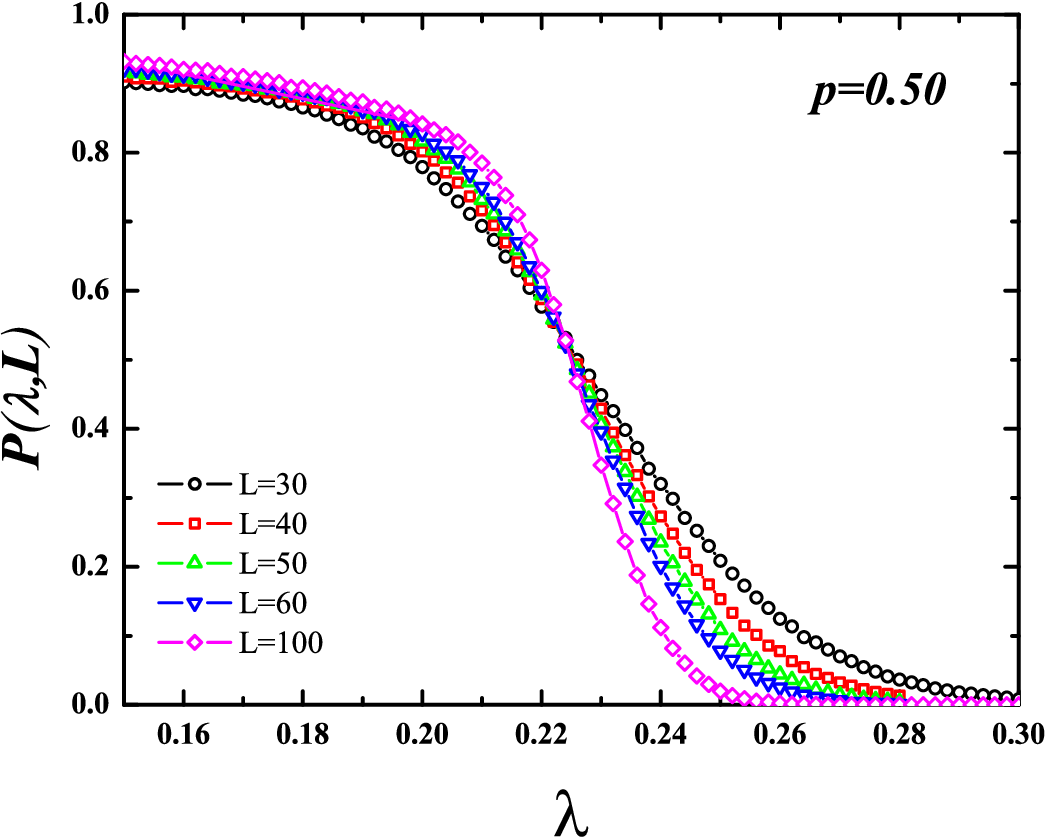}}
\qquad
\subfigure[]{\label{fig:02b}\includegraphics[scale=0.40, angle=0]{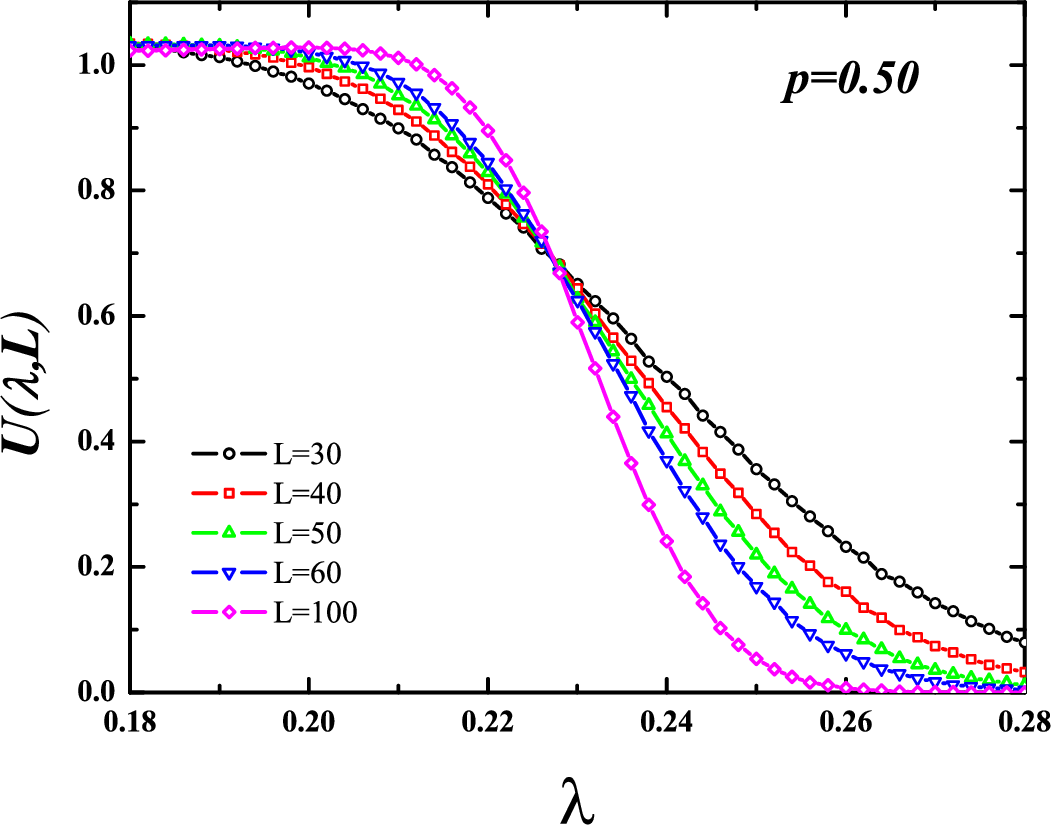}} 
\subfigure[]{\label{fig:02c}\includegraphics[scale=0.40, angle=0]{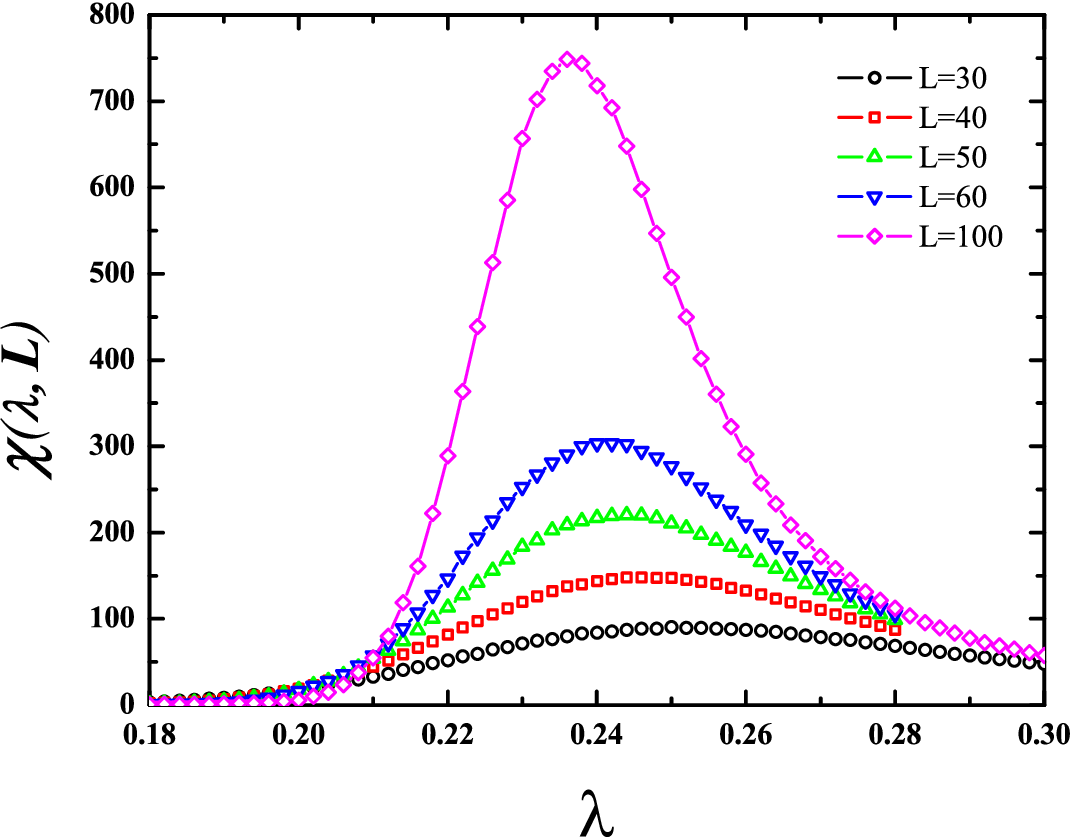}} 
\end{minipage}
\caption{Static quantities for the stochastic SIR model on square lattices with first- and second-neighbor interactions by assuming a dilution rate of $p=0.50$. Panels (a), (b), and (c) display the order parameter $P$, Binder cumulant $U$, and susceptibility $\chi$ as a function of the recovery rate $\lambda$, respectively. From the Binder cumulant crossing, we can estimate the epidemic threshold at $\lambda _{c}=0.226(7)$ for this case. Each data point represents averages over $10^5$ independent realizations.}\label{fig:02}
\end{figure*}

\begin{figure}[!t]
	\centering
		\includegraphics[scale=0.45, angle=0]{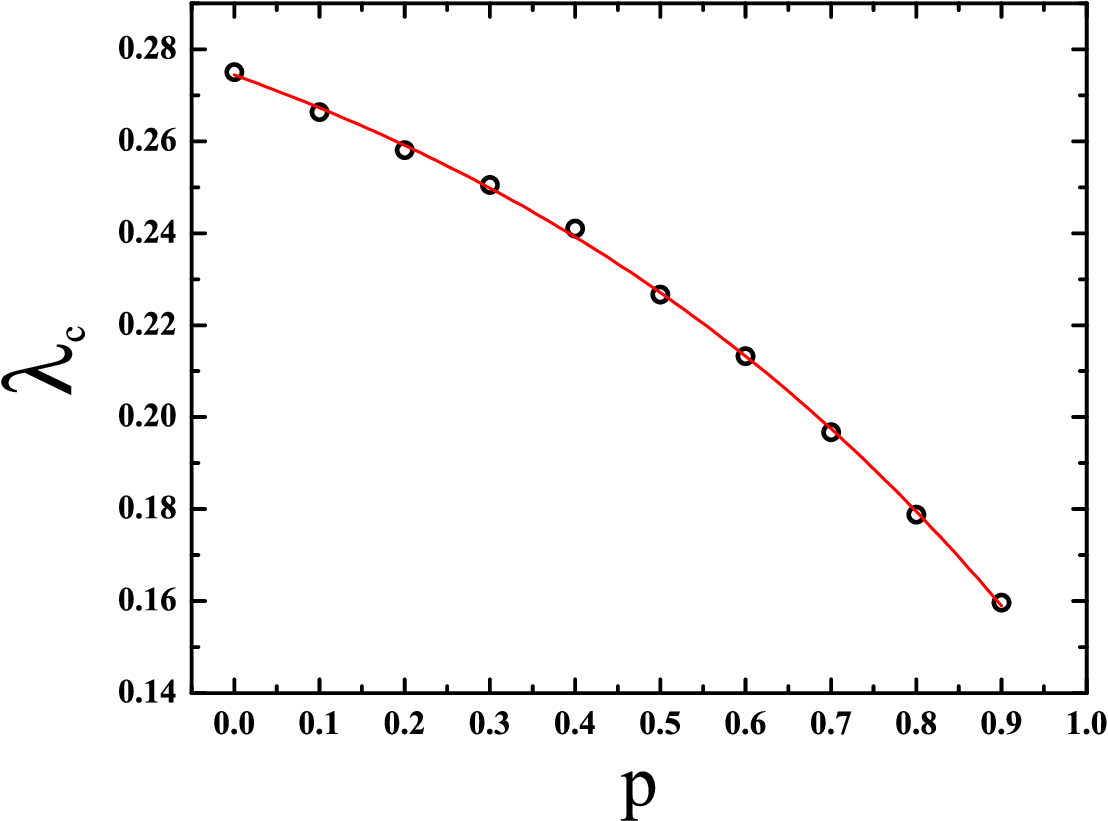}
		\caption{Plot for the critical threshold point for all considered $p$ cases. The red curve represents the non-linear curve fitting according to $\lambda_{c}=\lambda_{0}-a\exp(b\,p)$, being $\lambda_{0}=0.325$, $a=0.05$ and $b=1.32$.}\label{fig:03}
\end{figure}

\begin{figure*}[!t]
\centering
\begin{minipage}[t]{1.0\linewidth}
\centering
\subfigure[]{\label{fig:04a}\includegraphics[scale=0.40, angle=0]{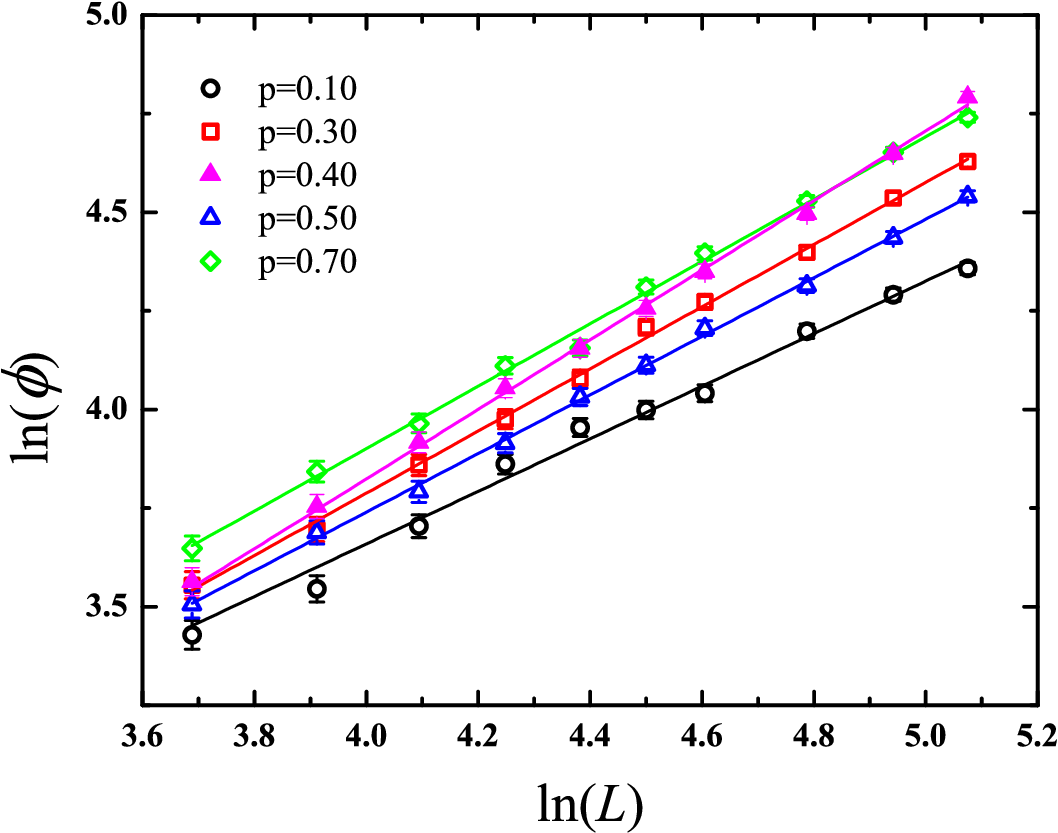}}
\qquad
\subfigure[]{\label{fig:04b}\includegraphics[scale=0.40, angle=0]{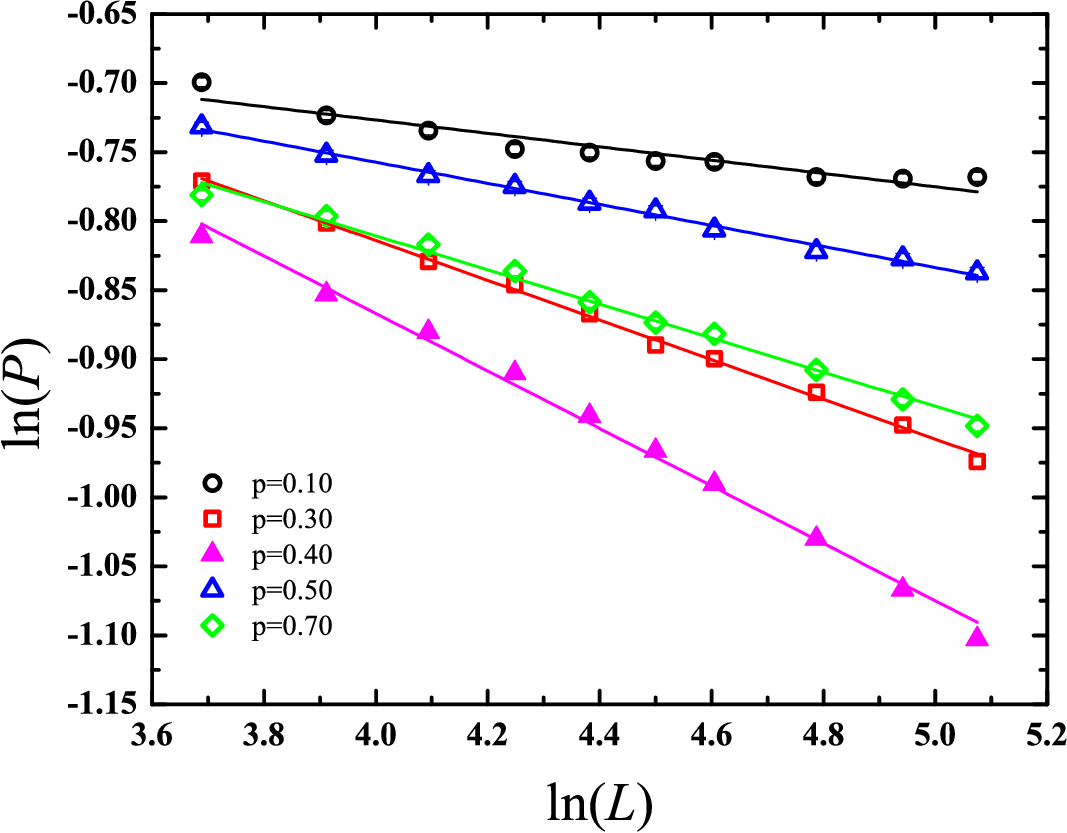}} 
\subfigure[]{\label{fig:04c}\includegraphics[scale=0.40, angle=0]{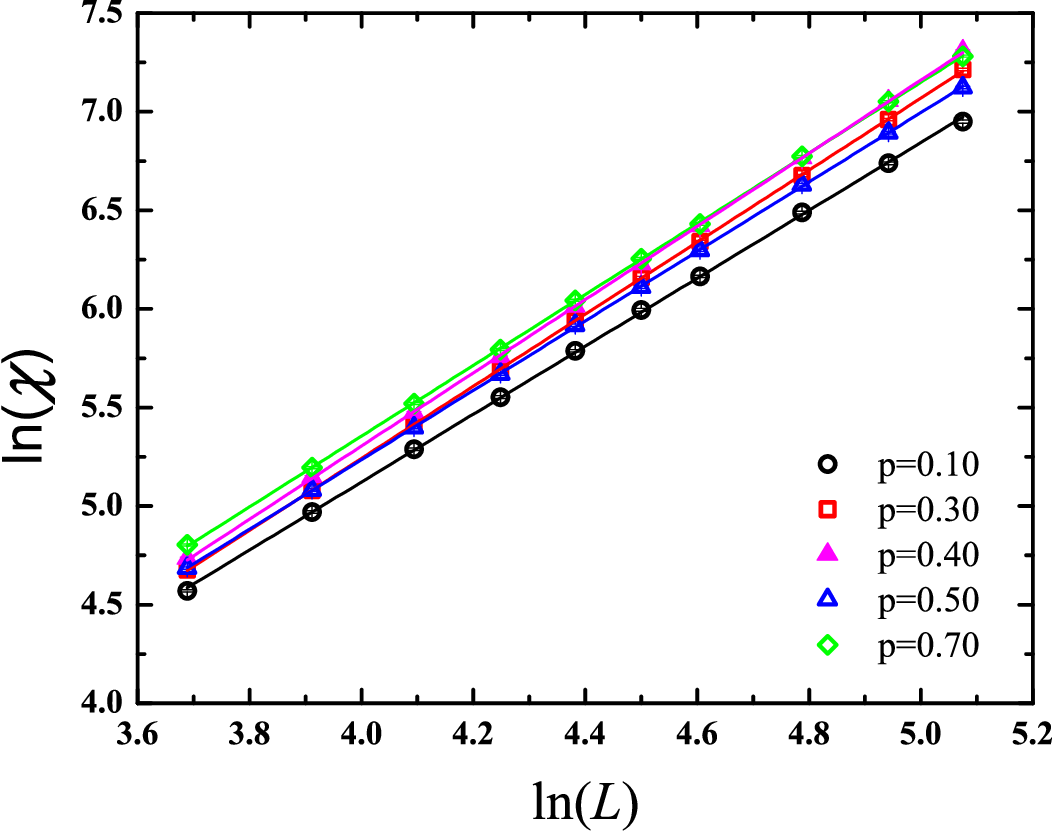}} 
\end{minipage}
\caption{Panels (a), (b), and (c) display the log-log plot of the quantities $\phi$, $P$, and $\chi$, respectively, for several bond-dilution cases ($p$) calculated at $\lambda_{c}$ as a function of the linear size of the system $L$. Straight lines represent the best linear fit to the corresponding data.}\label{fig:01}
\end{figure*}

In the critical region, the cluster size distribution should obey a power-law scaling \cite{Stauffer1979,Hoshen1979,Souza2011} as
\begin{equation} \label{eq:10}
	n_p (s) = s^{ - \tau } F[s^\alpha\; (\lambda  - \lambda _c )],
\end{equation}
where $\lambda _c$ is the epidemic threshold and $F$ is a scaling function. Likewise, the remaining quantities also adhere to scaling relations as prescribed by classical percolation theory.
\begin{align}
P &= L^{ - \beta /\nu \;} \tilde P(L^{1/\nu } {\kern 1pt} |\lambda  - \lambda _c |), \label{eq:11}\\
\chi  &= L^{\gamma /\nu \;} \tilde \chi (L^{1/\nu } {\kern 1pt} |\lambda  - \lambda _c |),  \label{eq:12}\\
< S' > & = L^{\gamma /\nu \;} \tilde S'(L^{1/\nu } {\kern 1pt} |\lambda  - \lambda _c |), \label{eq:13}\\
< M' > & = L^{(\beta  + 2\gamma )/\nu \;} \tilde M'(L^{1/\nu } {\kern 1pt} |\lambda  - \lambda _c |). \label{eq:14}
\end{align}
The reciprocal correlation-length exponents $1/\nu$ can be obtained by calculating the modulus of the logarithmic derivative of $P$ at the critical threshold point $\lambda_{C}$
\begin{equation} \label{eq:17}
	\phi \equiv \bigg |\frac{d}{{d\lambda }}\ln (P)\bigg |_{\lambda  = \lambda _c },
\end{equation}
where the derivative of the function $f\equiv \ln(P)$ was evaluated numerically by using a finite central difference scheme in the form
\begin{equation} \label{eq:18}
	\frac{{df}}{{d\lambda }} \simeq \frac{1}{{2\,\Delta \lambda}}\left( {f(\lambda  + \Delta \lambda) - f(\lambda  - \Delta \lambda)} \right),
\end{equation}
which has an truncation error of the order of $\mathscr{O}(\Delta \lambda^2)$. The error function $\delta f'$ of $df/d\lambda$ was obtained via error propagation from the uncertainties in the values of $f$ ($\delta f$), being expressed by
\begin{equation} \label{eq:19}
	\delta f'  = \frac{1}{{2\,\Delta \lambda}}\sqrt {(\delta f(\lambda  + \Delta \lambda))^2  + (\delta f(\lambda  - \Delta \lambda))^2 }.
\end{equation}
In our computations $\Delta \lambda=2.0 \times 10^{-3}$. Close to $\lambda_{c}$, the quantity $\phi$ obeys a power-law scaling as
\begin{equation} \label{eq:20}
\phi  = L^{1/\nu } \tilde \phi (L^{1/\nu } |\lambda  - \lambda _c |),
\end{equation}
where $\tilde \phi$ is also a scaling function. Moreover, we can define a universal quantity $U$ in which the scaling dependencies cancel out by combining Eqs.~(\ref{eq:11}), (\ref{eq:13}), and (\ref{eq:14}) in the following way \cite{Souza2011}
\begin{equation} \label{eq:15}
	U = P\frac{{ < M' > }}{{ < S' > ^2 }},
\end{equation}
being analogous to the Binder cumulant for ferromagnetic spin model \cite{Selke2005,Ferraz2015}, and obeying also a scaling relation
\begin{equation} \label{eq:16}
	U = \tilde U\; (L^{1/\nu } {\kern 1pt} |\lambda  - \lambda _c |).
\end{equation}
The crossing point of the $U$ curves for different lattice sizes allow us to estimate the epidemic threshold $\lambda _c$. Once $\lambda _c$ is determined, conducting finite-size scaling (FSS) analyses on the observables $P$, $\chi$, and $\phi$ using Eqs.~(\ref{eq:11}), (\ref{eq:12}), and (\ref{eq:20}) respectively, enables the calculation of the corresponding critical exponent ratios $\beta /\nu$, $\gamma /\nu$, and $1/\nu$. These ratios serve as indicators of the leading critical exponents $\beta$, $\gamma$, and $\nu$, which collectively characterize the universality class of the model.

The bond-diluted lattices utilized in this study were derived from regular square lattices with free boundary conditions. Initially, we begin with a standard square lattice where nodes are interconnected with their first and second nearest neighbors through both outgoing and incoming links. Subsequently, with a probability of $p$, we disconnect a selected site from its respective nearest neighbors, ensuring at least one connection is retained to prevent isolated sites. This process is iterated for each site, resulting in a new lattice characterized by a node density of $q=1-p$ with connections to both first and second nearest neighbors. Such networks emulate a more realistic population, showcasing a diverse array of interconnected relationships among individuals. The special case arises when $p=0$, corresponding to pure lattices where all nodes maintain connectivity with their first and second neighbors. This scenario was explored in detail in Ref.~\cite{Ferraz2021}, and it was found that this case belongs to the same universality class as that of 2D direct percolation.
 It is important to note that only cases in which $p \leq 0.70$ have been considered in the present study, as increasing the bond dilution level beyond that could severally impact the cardinality of the generated lattices.

Spanning clusters emerge when the cluster of recovered individuals extends across the entire lattice, connecting opposite edges. This occurs when the cloud spans from one side to the other. We generated over $10^{5}$ spanning clusters for each $\lambda$ value considered to ensure accurate averaging of the relevant quantities.  We deal with several lattice sizes, ranging from $N=400$ up to $N=25600$. 

\section{\label{sec:r} Results and Discussion}

In this section, we show our numerical results of the stochastic SIR model coupled to bond-diluted lattices. In order to determine both the critical region and the order of the phase transition in this model on these lattices, we calculated the order parameter $P$, Binder cumulant $U$, and susceptibility $\chi$ for several $p$ cases in a wide range of the parameter $\lambda$. These quantities were averaged over $10^{5}$ different dynamic realizations of the SIR model. After identifying the critical region, we employed FSS analysis to obtain reliable estimates of the critical $\lambda$ and the leading critical exponents. For each $p$ case we consider different lattice configurations upon taking the disorder averages.

In Figs.~\ref{fig:01a}, \ref{fig:01b} and \ref{fig:01c} are shown the order parameter, Binder cumulant, and susceptibility as a function of the recovery rate $\lambda$ for the case $p=0.20$, respectively. While Figs.~\ref{fig:02a}, \ref{fig:02b} and \ref{fig:02c} show the same quantities for the case $p=0.50$. As one can see from Figs.~\ref{fig:01a} and  ~\ref{fig:02a}, a typical second-order phase transition take place for these cases in which a typical sigmoid-shaped curve occurs. Similarly, the same conclusion can be drawn for the remaining treated cases ($p\leq 0.70$). From Binder cumulant crossings, we can estimate the corresponding epidemic thresholds for each case. The critical thresholds were estimated with four significant figures. For the case $p=0.20$, we obtained $\lambda_{c}=0.176(6)$, while for the case $p=0.50$, we got $\lambda_{c}=0.226(7)$. Fig.~\ref{fig:03} displays the critical threshold point for all considered cases, including cases with high dilution degrees, namely $p=0.80$ and $p=0.90$. The red curve represents the best non-linear fit to the corresponding data. One can observe from Fig.~\ref{fig:03} that the critical point for dilution disorder monotonically diminishes with increasing $p$ value.

By taking the slope of the log-log plot of the quantity $\phi$ versus the linear size of the system $L$, we can estimate $1/\nu$ for each case. Fig.~\ref{fig:04a} shows the best linear fit to Eq.~(\ref{eq:20}) for several cases. Error bars were estimated by using Eq.~(\ref{eq:19}). We obtained 
$1/\nu=0.667\pm 0.022$ for $p=0.10$, $1/\nu=0.691\pm 0.019$ for $p=0.20$, $1/\nu=0.788\pm 0.011$ for $p=0.30$, $1/\nu=0.883\pm 0.013$ for $p=0.40$, $1/\nu=0.743\pm 0.009$ for $p=0.50$, $1/\nu=0.775\pm 0.018$ for $p=0.60$, and $1/\nu=0.790\pm 0.016$ for $p=0.70$. As can be seen, the values of $1/\nu$ for the cases $p=0.10$, $p=0.20$, $p=0.30$, and $p=0.40$ are several standard deviations away from the exact critical exponent ratio $1/\nu=3/4$ of 2D dynamical percolation. Meanwhile, for the remaining cases, the estimated values of $1/\nu$ are within a few standard deviations of that exact value. 

Similarly, a finite-size scaling analysis of the magnitudes of the order parameter $P$ and the susceptibility $\chi$ at $\lambda_{c}$ by using Eqs.~(\ref{eq:11}) and (\ref{eq:12}) yield, respectively, the critical exponent ratios $\beta /\nu$ and $\gamma /\nu$ for all cases. Figs.~\ref{fig:04b} and \ref{fig:04c} show the log-log plot of $P$ and $\chi$ (both calculated at $\lambda_{c}$) against $L$, respectively. The straight lines in those figures are the best linear fit to Eqs.~(\ref{eq:11}) and (\ref{eq:12}), respectively. The estimates for $\beta /\nu$ and $\gamma /\nu$ for all cases are shown in Table \ref{tab:1}.

Most of these values deviate significantly, by several standard deviations, from the exact critical exponent ratios of 2D dynamical percolation, specifically $\beta /\nu=5/48$ and $\gamma /\nu=43/24$. With the exception of the case where $p=0.60$, all other estimates strongly suggest that the SIR model on a bond-diluted lattice, in general, does not belong to the same universality class as 2D dynamic percolation. Estimates of the critical exponents ratios and critical threshold $\lambda_{c}$ for all cases are summarized and compared with the corresponding exact values form 2D dynamics percolation in Table \ref{tab:1}.

\section{\label{sec:c} Conclusions}

We conducted Monte Carlo simulations of the stochastic SIR model on bond-diluted square lattices to investigate the critical behavior exhibited by these systems. We estimated both the critical threshold and the leading critical exponent ratios for various cases. Our numerical analysis has revealed that quenched bond dilution disorder significantly influences the critical exponents of the model for most of the considered $p$ values. The simulation results clearly suggest that the present model belongs to a new universality class different from that of 2D dynamical percolation, depending on the chosen $p$ value. Additionally, we observed that the critical threshold in the stochastic SIR model monotonically decreases with increasing bond-dilution disorder.

This study has broad relevance to many areas, not only to the dynamics of infectious disease transmission but also to general diffusion processes, the propagation of damage in random networks, and provides insights into how perturbations in connectivity affect the overall behavior, resilience, and robustness of complex systems. Additionally, the findings outlined in this paper may contribute to a deeper understanding of how topological irregularities can affect the critical properties of interconnected complex systems.

\section{Conflict of Interest Statement}
The authors have no competing interests to declare that are relevant to the content of this article.

\section{\label{sec:ref} References}


\bibliographystyle{elsarticle-num-names}

\end{document}